# Optical amplification and pulse interleaving for low noise photonic microwave generation

Franklyn Quinlan,[1,*] Fred N. Baynes,[1] Tara M. Fortier,[1] Qiugui Zhou,[2] Allen Cross,[2]
Joe C. Campbell,[2] and Scott A. Diddams[1]

[1]National Institute of Standards and Technology, Time and Frequency Division, Boulder, CO 80305
[2]Department of Computer and Electrical Engineering, University of Virginia, Charlottesville, Virginia 22904
*Corresponding author: fquinlan@boulder.nist.gov

We investigate the impact of pulse interleaving and optical amplification on the spectral purity of microwave signals generated by photodetecting the pulsed output of an Er:fiber-based optical frequency comb. It is shown that the microwave phase noise floor can be extremely sensitive to delay length errors in the interleaver, and the contribution of the quantum noise from optical amplification to the phase noise can be reduced ~10 dB for short pulse detection. We exploit optical amplification, in conjunction with high power handling modified uni-traveling carrier photodetectors, to generate a phase noise floor on a 10 GHz carrier of -175 dBc/Hz, the lowest ever demonstrated in the photodetection of a mode-locked fiber laser. At all offset frequencies, the photodetected 10 GHz phase noise performance is comparable to or better than the lowest phase noise results yet demonstrated with stabilized Ti:sapphire frequency combs.

Generating and distributing low phase noise microwave signals continues to be compelling for a number of scientific applications such as microwave atomic clocks [1], synchronization at large-scale facilities [2], and radar systems [3]. Optical frequency division (OFD), where a stable optical frequency is coherently divided to the microwave domain by an optical frequency comb, has recently emerged as a technique to generate microwave signals with extremely high spectral purity [4]. With this technique, 10 GHz signals have been produced that exhibit absolute phase noise below -100 dBc/Hz at 1 Hz offset, more than 40 dB below that of conventional room temperature microwave oscillators [5].

The lowest OFD absolute phase noise performance to date has been achieved with Ti:sapphire comb-based systems [6]. Er:fiber-based OFD systems are also of interest, due to their advantageous center wavelength for pulse distribution via optical fiber, the lower cost and power requirements for fiber lasers, and the amenability to a compact, mobile microwave source. While the phase noise performance of Er:fiber-based OFD systems is on par with the best Ti:sapphire OFDs for offset frequencies below a few kHz, at higher offset frequencies Er:fiber OFDs suffer from higher noise of the Er:fiber comb, and higher detection noise floor [7, 8]. For example, Ti:sapphire OFDs have demonstrated phase noise as low as -179 dBc/Hz at 10 MHz offset from a 10 GHz carrier, whereas the best Er:fiber OFDs to date range from -145 to -160 dBc/Hz at this offset frequency. While photonic-microwave hybrid oscillator systems are available to lower the noise far from carrier [9, 10], here we investigate the limits of the purely photonic approach to low noise microwaves with an Er:fiber OFD, demonstrating phase noise at 1 MHz (10 MHz) offset of a 10 GHz carrier of -170 dBc/Hz (-175 dBc/Hz). Across the phase noise spectrum, the Er:fiber OFD phase noise is comparable to, and at some frequencies better than, Ti:sapphire OFDs, approaching the far from carrier phase noise levels of the best microwave sources [9]. As discussed below, important to achieving these noise levels has been an analysis of the impact of pulse interleaver delay errors and optical amplification on the microwave phase noise, as well as improved servo control of the Er:fiber comb.

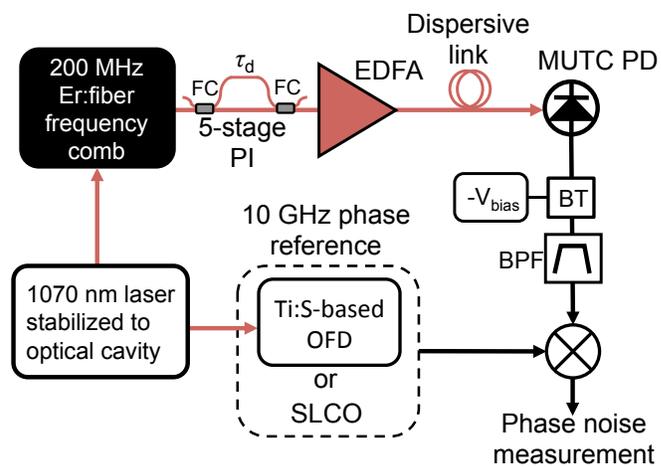

Fig. 1. Er:fiber-based OFD phase noise measurement setup. Only one stage of the 5-stage pulse interleaver is represented. BT, bias tee; PI, pulse interleaver; EDFA, erbium-doped fiber amplifier; SLCO, sapphire loaded cavity oscillator; BPF, 10 GHz bandpass filter; FC, fiber coupler. $\tau_d$ is the delay between the arms of the PI, a multiple of 100 ps for 10 GHz signal generation.

The Er:fiber OFD system and microwave phase noise measurement scheme are shown in Fig. 1. A cavity-stabilized laser at 1070 nm ($\nu_{opt}$ = 282 THz) is used as the optical frequency reference. The frequency comb source is an Er-doped fiber mode-locked laser, centered at 1550 nm with repetition rate $f_r$ = 208 MHz. A short intracavity free-space section includes waveplates and a polarization beam splitter to excite nonlinear polarization rotation mode-locking [7, 11]. The free-space section also includes a 1 cm long LiNbO$_3$ electro-optic phase modulator for cavity length stabilization. A standard f-2f interferometer is used for offset frequency ($f_o$) detection. Locking to $\nu_{opt}$ is accomplished by splitting off part of the supercontinuum light generated for the f-2f interferometer to create a heterodyne beat note $f_b = \nu_{opt} - (nf_r + f_o)$, where n is the comb mode number [7]. While low noise microwaves can be generated via independent locking of $f_o$ and $f_b$, lower noise may be achieved by electronically mixing the $f_o$ and $f_b$ signals to produce a single error signal $(f_b + f_o) = \nu_{opt} - nf_r$ [12]. Feedback of this error signal to the high-bandwidth (>1 MHz) intracavity electro-optic phase modulator stabilizes the repetition rate $f_r = [\nu_{opt} - (f_b + f_o)]/n$. Low bandwidth, large dynamic range cavity length control is accomplished with a piezoelectric transducer on an intracavity fiber coupler. Large drifts in $f_o$ are measured and compensated through a frequency-to-voltage converter controlling the pump laser current.

Phase noise measurements were performed on a 10 GHz carrier, corresponding to the 48th harmonic of $f_r$. When illuminated with the pulse train directly from the Er:fiber laser, photodetector saturation restricts the maximum power at 10 GHz to -5 dBm at 3 mA average photocurrent. This in turn limits the achievable signal-to-noise ratio (SNR), ultimately bounded by thermal (Johnson) noise at -169 dBc/Hz. Optical pulse interleaving, where an optical pulse train is split and recombined with a half-period delay, is a straightforward, and cascadable, method for increasing the microwave saturation power [8]. Four successive interleaver stages were built by fusion splicing 3 dB fiber couplers, increasing the pulse repetition rate to 3.33 GHz. A fifth stage created pairs of 100 ps-spaced pulses to further increase the saturation power at 10 GHz [13]. The detector used is a 40 μm diameter modified uni-traveling carrier photodetector (MUTC PD), designed for high power handling and high linearity [14]. The MUTC PD is flip-chip bonded onto an AlN substrate and contacted to a thermoelectric cooler for efficient heat removal. Such devices have demonstrated nearly 30 dBm of microwave power under sinusoidally-modulated illumination. With a five-stage pulse interleaver, the saturation power at 10 GHz increased to +23 dBm at 60 mA photocurrent, a 28 dB improvement and the highest microwave power produced from short pulse illumination of a photodetector of which we are aware.

For highest 10 GHz microwave power, the relative delay in each interleaver stage should be a multiple of 100 ps. For interleavers built from fiber couplers, the interleaver delays are not infinitely precise, reducing the power in the 10 GHz signal and the thermal floor-limited SNR. But as shown in Fig. 2, the microwave power is relatively insensitive to errors in the delay, and nearly full improvement in the thermal noise-limited phase noise is readily achieved [13]. At high photocurrent levels, the phase noise floor may not be limited by thermal noise, since shot noise will also impact the phase stability of the microwave signal. We have shown previously that for detection of a periodic train of ultrashort optical pulses, the shot noise resides primarily in the amplitude quadrature, reducing the impact on the microwave phase stability [15]. This phase noise reduction is quickly lost, however, when there are errors in the interleaver delays, as also shown in Fig. 2 for a single interleaver stage. Here we have used Eq. (20) of Ref. [16] to calculate the phase noise deviation for a train of 1 ps pulses with periodic shifts in the pulse timing due to the interleaver delay error. Clearly the shot noise-limited phase noise is quite sensitive to the interleaver delay, with an error of only 5 ps producing a 15 dB increase in the noise floor. This is contrasted with a decrease in 10 GHz power of less than 1 dB (and corresponding 1 dB increase in the thermal noise-limited floor) for the same error in the delay. The noise increase is due to the fact that the shot noise reduction relies not only on short optical pulses, but also on the exquisitely regular timing of an optical pulse train. When the timing regularity is disturbed by an interleaver delay error, the shot noise contribution to the microwave phase noise increases. Measurements using a two-stage free-space interleaver and a 5 GHz carrier confirmed this behavior [17]. Further measurements were conducted in conjunction with optical amplifier testing, as reported below.

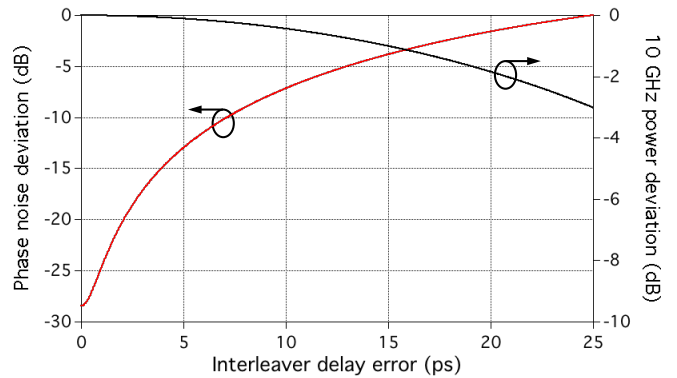

Fig. 2. Calculated 10 GHz signal power and shot noise-limited phase noise as a function of delay error in a single-stage optical pulse interleaver. The interleaver delay error is measured against any multiple of 100 ps. Dispersion differences between different paths through the interleaver are considered negligible. The phase noise deviation is referenced against shot noise contributing equally to amplitude and phase quadratures.

At the shot noise limit, Fig. 2 implies delay errors must be limited to a few picoseconds in order to prevent a significant phase noise floor increase. For a pulse interleaver built in fiber, this corresponds to a precision of the delay length of a few hundred micrometers. By polishing the fiber ends to fine-tune the delay length before splicing, we were able to build five fiberized interleaver stages with delay errors of 1.25 ps, 0.54 ps, 0.2 ps, 0.4 ps, and 1 ps. Pulse interleavers will also produce amplitude imbalance as pulses go through different optical paths. However, our analysis indicates the phase noise floor is largely insensitive to amplitude imbalances of this kind.

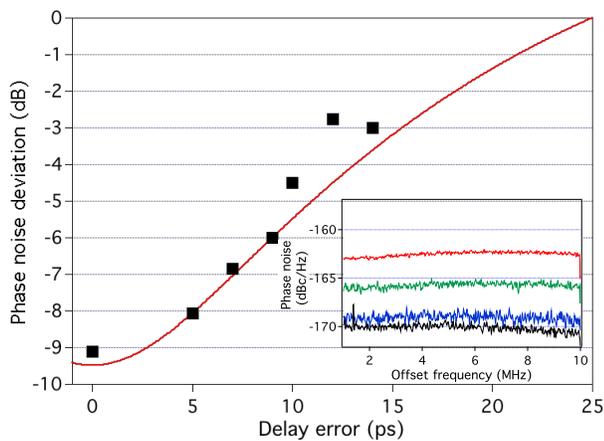

Fig. 3. EDFA-limited phase noise deviation as a function of interleaver delay error, compared to our analytical model (red trace). Inset: Phase noise at delay errors of 14 ps (red), 9 ps (green), 5 ps (blue), and minimized delay error (black).

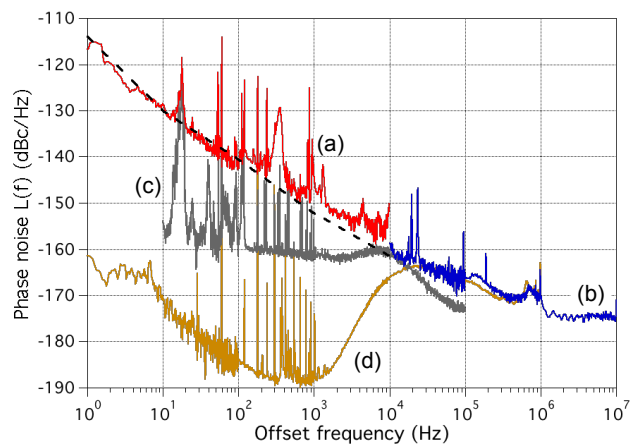

Fig. 4. Single-sideband phase noise on a 10 GHz carrier. (a) Er:fiber OFD compared against Ti:sapphire OFD, both locked to the same optical reference. (b) Er:fiber OFD compared to a sapphire loaded cavity oscillator. (c) Predicted RIN contribution to the phase noise through amplitude-to-phase conversion of the MUTC PD (d) Residual (in-loop) phase noise of the lock of the Er:fiber comb to the optical reference laser, scaled to a 10 GHz carrier. Dashed line represents the residual noise of two Ti:sapphire-illuminated MUTC PDs.

In order to take full advantage of the lower noise floor achievable with higher microwave saturation power, the optical power on the detector must also increase. Such power is not always available directly from the mode-locked laser. Here we investigate the impact of optical amplification on the microwave phase noise floor. Quantum noise added during optical amplification induces optical pulse-to-pulse timing jitter and pulse energy fluctuations that manifest themselves as noise in the photocurrent. The photocurrent noise from amplification is traditionally separated into spontaneous-spontaneous beat noise and signal-spontaneous beat noise [18]. In addition to these, Gordon-Haus jitter [19], where a random shift of the center frequency of each pulse couples with dispersion in the fiber link to produce timing jitter, will also contribute to the microwave phase noise. Despite the added noise, optical amplification may improve the phase noise performance. This is due to the fact that, as with shot noise, signal-spontaneous beat noise should display an imbalance between amplitude and phase quadratures, leading to the same optical pulse width-dependent contribution to the microwave phase noise [16]. Thus even though signal-spontaneous beat noise dominates the photocurrent noise, its contribution to the microwave phase noise can be very small. Moreover, just as with shot noise, errors in the interleaver delays should impact the signal-spontaneous beat noise contribution to the microwave phase noise. To test this, the pulses were sent through a four-stage fiberized pulse interleaver followed by a tunable free-space stage before passing through an erbium-doped fiber amplifier (EDFA). The delay of the final stage was varied, and the phase noise measured for offset frequencies of 1 MHz to 10 MHz. The total photocurrent SNR was also measured for each delay, so as to compare to the phase noise level. In order to more clearly observe the impact of pulse interleaver delay errors on the phase noise, the EDFA used for these experiments had a large noise figure, measured at ~10 dB. Phase noise data for selected delay errors are shown in the inset of Fig. 3. A noise increase is apparent with only a 5 ps error in the interleaver delay. With the delay error minimized, a phase noise floor of -170 dBc/Hz is achievable, despite large noise figure of the optical amplifier. Fig. 3 also compares the measured and predicted phase noise deviations as a function of delay error. The noise prediction is calculated by applying Eq. (20) of Ref. [16], incorporating the measured delay errors from the four fiberized interleaver stages. The phase noise deviation is defined as the noise level compared to one-half the measured SNR of the photocurrent, that is, the deviation from $S_n/(2P_{10G})$. Here $S_n$ is the noise power spectral density measured directly from the MUTC PD at a frequency near 10 GHz, and $P_{10G}$ is the power in the 10 GHz carrier. This gives the phase noise reduction compared to the optical amplifier contributing noise equally to amplitude and phase quadratures of the microwave signal. The data and the predicted noise level show reasonable agreement. When the delay error is minimized the phase noise floor is reduced ~9 dB. The only adjustable parameter to the model is the -9 dB floor at zero delay error. Possible origins of this floor are Gordon-Haus jitter or an unknown electronic noise source. Additional timing jitter due to randomness in the impulse response of the detector is not accounted for in the model, and its impact is currently under investigation.

To evaluate the Er:fiber OFD with best overall performance, the final free-space interleaver stage was replaced with the fiberized stage, and a new EDFA was built with a lower noise figure of ~5 dB with input power of 9 mW. The chosen EDFA fiber has normal dispersion to reduce the dispersion in the link to the MUTC PD. Different optical paths through the pulse interleavers yielded only minor differences in the optical pulse widths, resulting in ~1.5 ps pulses on the MUTC PD. Amplitude-to-phase noise conversion was measured on the MUTC PD [20], and a minimum was found at 16 mA, a photocurrent level not achievable without optical amplification. All phase noise measurements were performed at this photocurrent. The power after the 10 GHz microwave bandpass filter was 7 dBm. Phase noise measurements of the optimized Er:fiber OFD are shown in Fig. 4. For offset frequencies ranging from 1 Hz to 10 kHz, a Ti:sapphire OFD locked to the same 1070 nm stabilized laser was used as the microwave phase reference. This essentially removed the phase noise contribution from the optical frequency reference, allowing for direct observation of the noise added by the frequency combs, pulse interleaver, optical amplifier, and photodetection. Noise that is known to originate in the Ti:sapphire OFD are the peak near 350 Hz, and the noise from 1 kHz to 10 kHz. Below 1 kHz, the quiescent noise level is comparable to previously measured residual noise of Ti:sapphire-illuminated MUTC PDs [6]. For offset frequencies 10 kHz and beyond, phase noise was measured by comparing to a 10 GHz sapphire loaded cavity oscillator whose phase noise at 10 kHz offset is -160 dBc/Hz, dropping to -190 dBc/Hz at 1 MHz. From 10 kHz to 20 kHz, the projected impact of the Er:fiber OFD's relative intensity noise, estimated through the MUTC PD's amplitude-to-phase conversion factor at our operating current, overlaps with the measured phase noise, and most likely dominates the phase noise performance of the Er:fiber OFD at these frequencies. From 20 kHz to ~1 MHz, the noise follows the residual noise in the Er:fiber comb lock to the optical reference. This noise is >20 dB lower than the results we obtain by independently locking $f_o$ and $f_b$. The slight increase in noise at ~150 kHz is known to originate in the 1070 nm reference laser. Beyond 1 MHz, the phase noise floor reaches -175 dBc/Hz. This floor is 12 dB below half the measured photocurrent SNR, indicating the optical amplifier noise largely resides in the amplitude quadrant of the 10 GHz carrier. Without the optical amplifier, the 10 GHz power after the microwave bandpass filter is -3.5 dBm, with a corresponding phase noise floor at the thermal noise limit of -173.5 dBc/Hz. Thus due to the small impact on microwave phase noise, not only can a phase noise increase be avoided with optical amplification, but an improvement over what was possible without optical amplification can be achieved.

In conclusion, when employing optical pulse interleaving for microwave signal generation, minimizing the errors in the interleaver delays is required to maintain the lowest shot-noise limited microwave phase noise floors. Also, despite their added noise, optical amplifiers contribute very little to the microwave phase noise when detecting ultrashort optical pulses, and in some cases can lower the microwave phase noise floor. While applicable in the detection of any train of ultrashort optical pulses, here we utilize carefully constructed pulse interleavers and optical amplification to generate 10 GHz signal from an Er:fiber OFD, yielding state-of-the-art phase noise performance of -175 dBc/Hz at 10 MHz offset from a 10 GHz carrier.

We thank William Loh and Pascal Del'Haye for helpful comments on this manuscript. This research was support by NIST and the DARPA PULSE program. It is a contribution of an agency of the US government and is not subject to copyright in the USA.